\documentclass[preprint,prd,showpacs,amssymb,floatfix]{revtex4}
\usepackage{graphicx}
\usepackage[latin1]{inputenc}
\usepackage{amsmath}
\usepackage{amsfonts}
\usepackage{amssymb}
\usepackage{bm}

\begin{document}

\title{\bf Spectral Line Broadening and Angular Blurring due to Spacetime 
Geometry Fluctuations}
\author{R. T. Thompson}
\email{robert@cosmos.phy.tufts.edu} 
\author{L.H. Ford}
\email{ford@cosmos.phy.tufts.edu}
\affiliation{Institute of Cosmology \\ 
Department of Physics and Astronomy \\
Tufts University, Medford, MA 02155}

\begin{abstract}
We treat two possible phenomenological effects of quantum fluctuations 
of spacetime geometry: 
spectral line broadening and angular blurring of the image of a distance 
source. A geometrical construction will be used to express both effects 
in terms of the Riemann tensor correlation function. We  apply
the resulting expressions to study some explicit examples in which the
fluctuations arise from a bath of gravitons in either a squeezed state
or a thermal state. In the case of a squeezed state, one has two limits
of interest: a coherent state which exhibits classical time variation
but no fluctuations, and a squeezed vacuum state, in which the fluctuations
are maximized.  
\end{abstract}


\maketitle

\baselineskip=14pt

\section{Introduction}
Spacetime fluctuations are expected to be a generic feature of a
theory which combines quantum theory and gravitation.  While  there is 
still no  complete quantum theory of gravity, it is possible to
investigate some of the characteristics expected of fluctuating spacetime
geometries. One can roughly classify these fluctuations as being either
active or passive.
Active fluctuations arise from fluctuations of the dynamical degrees
of freedom of gravity itself, that is, from the quantization of gravity.
  Passive fluctuations arise from fluctuations in the stress tensor due to
quantum matter fields. In general, one expects both types of fluctuations to 
be present. One approach to the study of fluctuating spacetimes is
stochastic gravity~\cite{Stochastic}.  
More generally, there has been considerable
activity in recent years in the area of quantum gravity phenomenology,
which seeks to find observational signatures of the quantum nature of 
spacetime~\cite{Moffat,JR94,GLF,HS97,EMN00,NvD00,AC00,NCvD03,AC04,DHR04,
Borgman}.  
There has also been considerable attention given to the effects of classical
stochastic gravitational fields~\cite{Zipoy,Sachs,Kaufmann,BM71,BKPN90}
 and to scattering of probe particles by gravitons in
an S-matrix approach~\cite{Weber,DeWitt}.  

Spacetime geometry fluctuations should in principle produce observable
effects on test particles, such as light rays. Several effects of
fluctuating gravitational fields on light propagation have been discussed
by previous authors. For example, Sachs and Wolfe~\cite{Sachs} treated the 
scattering of cosmic microwave photons from cosmological density perturbations.
Zipoy~\cite{Zipoy}  argued that there will be apparent luminosity variations
in a source seen through a stochastic gravitational field. More recently,
this effect has been treated~\cite{Borgman} using a Langevin form of the 
Raychaudhuri equation~\cite{Moffat}. In the latter approach, luminosity
fluctuations are a signature of passive spacetime geometry fluctuations
caused by quantum stress tensor fluctuations. Zipoy~\cite{Zipoy} also
examined changes in angular position due to a stochastic gravitational field.
Kaufman~\cite{Kaufmann} considered redshift fluctuations from a bath
of gravity waves. If one goes beyond the geometric optics approximation
to consider electromagnetic waves propagating in a stochastic gravitational
field, there will be phase shift variations~\cite{BM71,BKPN90,HS97}.

The purpose of the present paper is to analyze in more detail two particular
signatures of spacetime geometry fluctuations, redshift fluctuations
and angular blurring of images.  We derive  simple expressions 
involving the Riemann tensor correlation function  and use these 
to examine fluctuations in
redshift and angular position of a source. We will illustrate this approach
using the cases of gravitons in a squeezed  state and in a thermal state, 
and calculate the broadening of spectral lines and an angular blurring of 
an object viewed through  a region of spacetime filled with gravitons.
This will serve as a simple example of spacetime undergoing active
quantum fluctuations.

This paper is organized as follows:  In Sect.~\ref{sec:rab} we provide a
detailed derivation of the equations for redshift fluctuations and
angular blurring given in Ref.~\cite{peyresq10}.  The derivation is
independent of the source of fluctuations, being based on the
Riemann tensor correlation function.  In Sect.~\ref{sec;squeeze} we assume
that gravitons in a general squeezed state act as the source of
fluctuations and are described by the correlation function, and we then
specialize the results from Sect.~\ref{sec:rab} to this model. Several
examples are calculated for the variance of both redshift and
angular position, and are compared with the expected classical time
variation of these quantities.  We then express these results in
terms of the energy density of the gravity wave and make an
order of magnitude estimate for the size of the effect. 
In Sect.~\ref{sec:thermal} we assume a
thermal bath of gravitons as the source of fluctuations and again
specialize the results of Section~\ref{sec:rab} to this model.  The results of
the paper are summarized and discussed in Sect.~\ref{sec:final}. A brief review
of the pertinent information regarding squeezed quantum states
required for this paper can be found in the appendix. Units in which
$G = \hbar = c =1$ will be used, where $G$ is Newton's constant.

\section{Redshift and Angular Blurring in Linearized Gravity}
\label{sec:rab}

\subsection{Linearized Redshift}

\begin{figure}
  \centering
  \scalebox{.5}{\includegraphics{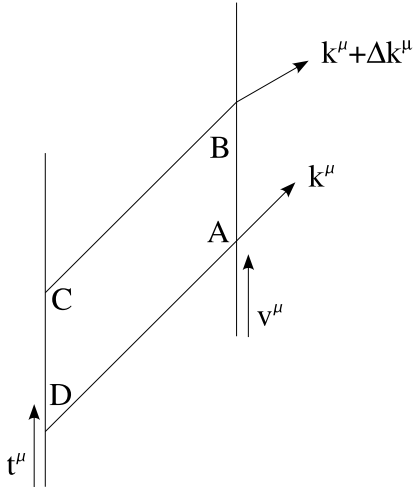}}\\
  \caption{A source moves along a worldline with tangent $t^{\mu}$
  while a detector a proper distance $s$ away moves along a worldline
  with tangent $v^{\mu}$. The source emits a ray at point D which has
  tangent $k^{\mu}(\lambda=0)$ at point D and tangent $k^{\mu}(\lambda_0)$ 
  at A.
  Parallel propagation of $k^{\mu}$
  around ABCD results in a slightly rotated vector $k^{\mu} + \Delta k^{\mu}$.
  The closed path ABCD encloses the spacetime region of interest.}
  \label{spacetime}
\end{figure}

Let $t^{\mu}$ be the 4-velocity of a source and $v^{\mu}$ the
4-velocity of a detector at the events of emission and absorption,
respectively, and let $k^{\mu}$ be tangent to the null geodesic
connecting them (see Fig.~\ref{spacetime}, e.g.,\ path $DA$).  Let
$\lambda$ be an affine parameter that runs from $\lambda =0$
(emission) to $\lambda=\lambda_0$ (absorption). Since $k^{\mu}$ is
tangent to a geodesic,
\begin{equation} \label{geodesiceqn}
 \frac{dk^{\mu}}{d\lambda} + \Gamma^{\mu}_{\alpha
 \beta}k^{\alpha}k^{\beta} = 0
\end{equation}
and so
\begin{equation}
 k^{\mu}(\lambda)=
k^{\mu}(0) - \int_0^{\lambda} d\lambda' \, \Gamma^{\mu}_{\alpha
 \beta}(\lambda')k^{\alpha}(\lambda')k^{\beta}(\lambda').
\end{equation}
This is an integral equation for $k^{\mu}(\lambda)$, but in the
linearized theory, we let $k^{\mu}(\lambda') \approx k^{\mu}(0)$ on
the right hand side.  So at the point of detection,
\begin{equation}
 k^{\mu}(\lambda_0)=k^{\mu}(0) - \int_0^{\lambda_0} d\lambda' \, 
\Gamma^{\mu}_{\alpha
 \beta}(\lambda')k^{\alpha}(0)k^{\beta}(0).
\end{equation}
The frequency of emission is proportional to $-k^{\mu}t_{\mu}$. The
proportionality constant depends on the choice of affine parameter.
If we choose the affine parameter such that, for example, $k^{\mu} =
(\omega,\omega,0,0)$ in the rest frame of the source, 
then $\omega = -t_{\mu}k^{\mu}.$ Since Eq.
(\ref{geodesiceqn}) is scale invariant, we may also select an affine
parametrization such that $k^{\mu} = (1,1,0,0).$ This amounts to
choosing the affine parameter to coincide with the source's proper
time at the evaluation point. We adopt the latter parametrization of
$\lambda$, in which case
 the ratio of the detected to the emitted frequency is
\begin{equation}
 \frac{\omega(\lambda_0)}{\omega_0} = -v_{\mu}k^{\mu}(\lambda_0) 
= -v_{\mu}k^{\mu}(0) + v_{\mu}
 \int_0^{\lambda_0} d\lambda' \, \Gamma^{\mu}_{\alpha
 \beta}(\lambda')k^{\alpha}(0)k^{\beta}(0) \, ,
\end{equation}
where $\omega_0 = \omega(0)$ is the emitted frequency.
The first term on the right hand side of this equation can
be written as
\begin{equation}
 -v_{\mu}k^{\mu}(0)= -t_{\mu}k^{\mu}(0) - (v_{\mu}-t_{\mu})k^{\mu}(0)
 = 1  - (v_{\mu}-t_{\mu})k^{\mu}(0).
\end{equation}
With this, the fractional change in detected frequency between
source and detector is
\begin{equation} \label{linearizedRF}
 \frac{\Delta \omega}{\omega_0} = \frac{\omega(\lambda_0)-\omega_0}{\omega_0} 
= -(v_{\mu}-t_{\mu})k^{\mu}(0) + v_{\mu}
 \int_0^{\lambda_0} d\lambda' \, \Gamma^{\mu}_{\alpha \beta}(\lambda')k^{\alpha}(0)k^{\beta}(0).
\end{equation}
The first term on the right hand side is a Doppler shift which may
not be zero even in flat space.  The second term is a linearized
gravitational redshift which depends on the intermediate geometry
between source and detector.  We will focus our attention on this
second term.

\subsection{Rate of Change of Redshift}

The next step is to examine the rate of change of the redshift.  We
examine the behavior of successive photons emitted from the source.
Our interest lies in the change in frequency due to the effects of
gravity rather than as a result of a variation of the output of the
source. This condition is enforced by requiring that the values of
$k^{\mu}$ at the start of any two null geodesics are related by
parallel transport along the worldline of the source. Thus we assume
that $\omega_0$ is constant between successive emissions and want
to find $\dot{\omega}(\lambda_0)$ at the detector due to changes in
spacetime geometry. Initially at the detector,
\begin{equation} \label{omega1}
 \omega(\tau_1,\lambda_0) = 
-v_{\mu}(\tau_1)k^{\mu}(\tau_1,\lambda_0)\, \omega_0
\end{equation}
is the frequency at the detector at proper time $\tau_1$. While
after a time $\Delta\tau$, at proper time $\tau_2$, the frequency at
the detector is
\begin{equation} \label{omega2}
 \omega(\tau_2,\lambda_0) =
 -v_{\mu}(\tau_2)k^{\mu}(\tau_2,\lambda_0)\, \omega_0 \, .
\end{equation}
To find $v^{\mu}(\tau_2)$, we first note that $v^{\mu}$ is tangent to
a geodesic:
\begin{equation}
 \frac{dv^{\mu}}{d\tau} + \Gamma^{\mu}_{\alpha
 \beta}v^{\alpha}v^{\beta} = 0 \, ,
\end{equation}
so
\begin{equation}
 v^{\mu}(\tau_2) - v^{\mu}(\tau_1) \approx \frac{dv^{\mu}}{d\tau}\Delta\tau
 \simeq -\Gamma^{\mu}_{\alpha\beta}v^{\alpha}v^{\beta}\Delta\tau.
\end{equation}
This depends on $\Gamma^{\mu}_{\alpha\beta}$ at the location of the
detector.  For simplicity, we assume $\Gamma^{\mu}_{\alpha\beta} =
0$ at the location of the detector.  This can be achieved by
assuming the detector is located in a flat region; more will be said
about this in the next subsection. Thus $v^{\mu}(\tau_2) =
v^{\mu}(\tau_1) = v^{\mu}.$

To find the change in $k^{\mu}$, we parallel transport the vector
around the closed path ABCD (see Fig.~\ref{spacetime}).  Recall that
if we parallel transport a vector around a closed path, the change
in the vector can be expressed as an integral of the Riemann tensor
over the area enclosed by the path.  For parallel transport around
an infinitesimal parallelogram, we have for an arbitrary vector
$V^{\mu}$
\begin{equation}
 V^{\mu} \rightarrow V^{\mu} + \delta V^{\mu}\, ,
\end{equation} 
where
\begin{equation}
 \delta V^{\mu} =
 -R^{\mu}_{\phantom{\mu}\alpha\nu\beta}V^{\alpha}t^{\nu}k^{\beta}\Delta\lambda\Delta\tau.
\end{equation}
Integrating over $\lambda$ and $\tau$ transports $V^{\mu}$ around a finite
parallelogram so that
\begin{equation}
 \Delta V^{\mu} = -\int d\tau \int d\lambda \,
 R^{\mu}_{\phantom{\mu}\alpha\nu\beta}V^{\alpha}t^{\nu}k^{\beta}.
\end{equation}
Since we assume the detector is located in a flat region, we can
further assume that $k^\mu$ may be trivially transported from point $A$ to 
point $B$, and we have 
\begin{equation}
k^{\mu}(\tau_1,\lambda_0) + \Delta k^{\mu}  = 
k^{\mu}(\tau_2,\lambda_0) \,.
\end{equation} 
Thus
\begin{equation}
 \Delta k^{\mu} = k^{\mu}(\tau_2,\lambda_0) - k^{\mu}(\tau_1,\lambda_0) =
 - \int_{\tau_1}^{\tau_2} d\tau \int_0^{\lambda_0} d\lambda \,
 R^{\mu}_{\phantom{\mu}\alpha\nu\beta}k^{\alpha}t^{\nu}k^{\beta}\,,
\end{equation}
and from Eqs.~(\ref{omega1}) and (\ref{omega2})
\begin{equation}
 \frac{\Delta\omega(\lambda_0)}{\omega_0} =
 \frac{\omega(\tau_2,\lambda_0)-\omega(\tau_1,\lambda_0)}{\omega_0}=
 -v_{\mu}\Delta k^{\mu} = v_{\mu}\int_{\tau_1}^{\tau_2}
 d\tau \int_0^{\lambda_0} d\lambda \,
 R^{\mu}_{\phantom{\mu}\alpha\nu\beta}k^{\alpha}t^{\nu}k^{\beta}.
\end{equation}
This equation can be thought of as relating the difference between
taking path DAB and path DCB. Taking limits in the previous equation
yields the rate of change of the fractional redshift in the
linearized theory:
\begin{equation}\label{RateOfChange}
 \frac{d}{d\tau}\left(\frac{\omega(\lambda_0)}{\omega_0}\right) 
= v_{\mu} \int_0^{\lambda_0}
 d\lambda \,
 R^{\mu}_{\phantom{\mu}\alpha\nu\beta}k^{\alpha}t^{\nu}k^{\beta}.
\end{equation}
An equivalent formula has been obtained by Braginsky and Menskii~\cite{BM71} 
using the geodesic deviation equation.

\subsection{Fluctuating Redshift}

Now suppose the Riemann tensor is subject to fluctuations - active,
passive, or both.  We first have
to specify $t^{\mu}$ and $v^{\mu}$ and how they behave under the
fluctuations. The simplest assumption is that $t^{\mu}$ and
$v^{\mu}$ do not fluctuate, and in the underlying flat geometry
$v^{\mu} = t^{\mu}$. Physically, this might be achieved via several
methods.  We have already mentioned that we can assume the detector
to be located in a flat region.  However, one could also consider
the source and the detector rigidly attached to one another by
non-gravitational forces.  The same effect might be achieved if the
source and detector are separately attached to platforms (e.g.\
planets) which are large enough that they travel on an average
geodesic in the mean spacetime.  This can happen if the spatial
average of the fluctuations over the platform is small. For our
purposes it is sufficient to assume the perturbation vanishes at
both the source and detector, e.g., a gravity wave passes between
source and detector, but far from either source or detector; so we can
assume the source and detector to be located in flat regions. The
fluctuations of the Riemann tensor are described by the correlation
function
\begin{equation}\label{CorFunction}
 C_{\alpha\beta\mu\nu\,\gamma\delta\rho\sigma}(x,x') = \langle
 R_{\alpha\beta\mu\nu}(x)
 R_{\gamma\delta\rho\sigma}(x')\rangle -
 \langle R_{\alpha\beta\mu\nu}(x)\rangle
 \langle R_{\gamma\delta\rho\sigma}(x')\rangle,
\end{equation}
where the indices $\alpha\beta\mu\nu$ refer to point $x$ while the
indices $\gamma\delta\rho\sigma$ refer to point $x'$.  With $v^{\mu}
= t^{\mu}$, we get from Eq.~(\ref{linearizedRF})
\begin{equation}
 \xi\equiv\frac{\Delta \omega}{\omega_0} = 
\frac{\omega(\lambda_0)-\omega_0}{\omega_0} = \int_0^{\lambda_0} d\lambda \,
 \Gamma^{\mu}_{\alpha\beta}(\lambda)k^{\alpha}(0)k^{\beta}(0)t_{\mu}.
\end{equation}
We let $\Gamma^{\mu}_{\alpha\beta}$ fluctuate with fixed $k^{\alpha}
= k^{\alpha}(0)$ and $t_{\mu},$ so that
\begin{equation}
 \langle\xi\rangle =
 \int_0^{\lambda_0}d\lambda \,
 \langle\Gamma^{\mu}_{\alpha\beta}(\lambda)
 \rangle k^{\alpha}k^{\beta}t_{\mu}.
\end{equation}
The integral in this equation is evaluated along a single line,
e.g., along the paths DA or CB. The equation may be
interpreted in the following way: Suppose the spacetime is subject
to quantum fluctuations. Then, given an ensemble of systems,
measurement of the fractional redshift along the same line will
yield different results.  This equation gives the expectation value
of those measurements.  Therefore, comparing the result of an
integration along path DA with that along path CB will give the
expectation of the difference of the fractional redshifts. Another,
more convenient way to obtain the same information is via 
Eq.~(\ref{RateOfChange}). Notice that
\begin{equation}
 \frac{d\xi}{d\tau} =
 \frac{d}{d\tau}\left(\frac{\omega(\lambda_0)-\omega_0}{\omega_0}\right) =
 \frac{d}{d\tau}\left(\frac{\omega(\lambda_0)}{\omega_0}\right)\,,
\end{equation}
 and thus from (\ref{RateOfChange}),
\begin{equation}
 \frac{d\xi}{d\tau} = \int_0^{\lambda_0}
 d\lambda \,
 R_{\alpha\beta\mu\nu}(\tau,\lambda)t^{\alpha}k^{\beta}t^{\mu}k^{\nu}.
\end{equation}
Integrating this expression yields
\begin{equation}
 \Delta\xi = \xi\bigr|_{\tau_2}-\xi\bigr|_{\tau_1} =
 \int_{\tau_1}^{\tau_2} d\tau \int_0^{\lambda_0}
 d\lambda \,
 R_{\alpha\beta\mu\nu}(\tau,\lambda)t^{\alpha}k^{\beta}t^{\mu}k^{\nu}.
\end{equation}
We now let $R_{\alpha\beta\mu\nu}(\tau,\lambda)$ fluctuate, and find
the expectation value $\langle\Delta\xi\rangle$ is
\begin{equation}
 \langle\Delta\xi\rangle = \int_{\tau_1}^{\tau_2}d\tau
 \int_0^{\lambda_0}d\lambda \, \langle
 R_{\alpha\beta\mu\nu}(\tau,\lambda)\rangle
 t^{\alpha}k^{\beta}t^{\mu}k^{\nu}.
\end{equation}

The expectation of the square, $\langle(\Delta\xi)^2\rangle$, is
related to the line broadening an observer will see.  Recall that
$\xi$ is the fractional change in frequency of an observed spectral
line, the fractional redshift.  Upon quantizing the
perturbation between source and detector we strictly must discuss
the fractional redshift along a given path as an ensemble average.
An observer, over some proper time interval $\Delta\tau$, collects
information on a distribution of fractional redshifts;
$\langle(\Delta\xi)^2\rangle$ is the squared  width
of this distribution.  The physical realization of this measurement
is a broadening of the observed spectral line.

However, there is a contribution to spectral line broadening due to
regular time dependent variations of the spacetime, for example as a
result of passing classical gravity waves. We thus characterize the
fluctuation of redshift about the classical time dependent variation
that arises from a nonzero expectation value of the Riemann tensor.
Using Eq.~(\ref{CorFunction}), we can express the variance of the
fractional redshift, $\delta\xi^2$, as
\begin{equation}
  \delta\xi^2 = \langle(\Delta\xi)^2\rangle -
  \langle\Delta\xi\rangle^2 =
  \int da \int da' \,
  C_{\alpha\beta\mu\nu\,\gamma\delta\rho\sigma}(x,x')
  t^{\alpha}k^{\beta}t^{\mu}k^{\nu}t^{\gamma}k^{\delta}t^{\rho}k^{\sigma}.
                    \label{eq:delta_xi}
\end{equation}
The integration over the spacetime region $\int da$ corresponds to
$\int_{\tau_1}^{\tau_2} d\tau \int_0^{\lambda_0} d\lambda$ and
similarly for $da'$.  This is an integration over the spacetime
region enclosed by the path ABCD in Fig.~\ref{spacetime}. The
point $x$ corresponds to the point $(\tau,\lambda)$ and similarly
for $x'$.

\subsection{Fluctuating Angular Position}

We can also relate the degree of angular blurring of the source
observed by the detector to the Riemann tensor correlation function.
Let $s^{\mu}$ be a unit spacelike vector in a direction orthogonal
to the direction of propagation of the null rays; thus
$s_{\mu}t^{\mu} = s_{\mu}k^{\mu}(\lambda=0) = 0$.  Then at the
observation point
\begin{equation}
 s_{\mu}k^{\mu}(\lambda_0) = \tan\Theta \approx \Theta,
\end{equation}
where $\Theta$ is an angle in the plane defined by the pair of
spacelike vectors $s^{\mu}$ and $n^{\mu} = k^{\mu}-t^{\mu}$ and is
assumed to be small, $|\Theta| \ll 1$. 
The angle $\Theta$ is the angular deviation in
the direction of $s^{\mu}$ of
the image of the source from its classical flat space position.  
A treatment similar to that shown for fractional redshift allows
us to express the change in angle, $\Delta\Theta$, in terms of an
integral of the Riemann tensor as
\begin{equation}
 \Delta\Theta = s_{\mu}\Delta k^{\mu} = \int da \, R_{\alpha\beta\mu\nu}
 s^{\alpha}k^{\beta}t^{\mu}k^{\nu}.
\end{equation}
A fluctuating spacetime results in an ensemble distribution of image
positions about the classical flat space position. Analogously to
the line broadening effects, the fluctuating angular position
manifests itself as a blurring of the source's image.
$\langle(\Delta\Theta)^2\rangle$ is therefore a measure of the
angular size of the image.  The variance of $\Delta\Theta$,
$\delta\Theta^2$, due to fluctuations in the Riemann tensor is
\begin{equation}
 \delta\Theta^2 = \langle(\Delta\Theta)^2\rangle -
 \langle\Delta\Theta\rangle^2 = \int da\int da' \,
 C_{\alpha\beta\mu\nu \, \gamma\delta\rho\sigma}(x,x')
 s^{\alpha}k^{\beta}t^{\mu}k^{\nu}s^{\gamma}k^{\delta}t^{\rho}k^{\sigma}.
                                    \label{eq:delta_theta}
\end{equation}

\subsection{Quantization of the Riemann Tensor} 

Our key results for the redshift fluctuations, Eq.~(\ref{eq:delta_xi}), 
and for the angular blurring, Eq.~(\ref{eq:delta_theta}), apply  both to 
active and passive fluctuations of spacetime geometry. We will present
some explicit examples for the case of active fluctuations. 
 We examine fluctuations produced by gravitons
occupying a squeezed state and then a thermal bath of gravitons in
the following two sections, respectively.  First we describe some of
the formalism relevant to both cases. A linearized quantum field
theory for gravity is used,  with the field operator expanded as
\begin{equation}
  \hat{h}_{\mu\nu} = 
\sum_{\ell, p} \left( A_{\mu\nu}\hat{a}_{\ell,p}
e^{i\ell_{\tau} x^{\tau}} +  h.c. \right)\, ,
\end{equation}
where $\ell^\mu$ is the wavevector and $p$ labels the polarization
of a mode with polarization tensor $A_{\mu\nu} = A_{\mu\nu}(\ell,p)$.
In this theory, the Riemann tensor operator is given by
\begin{equation} \label{MetricDerivatives}
 \hat{R}_{\alpha\beta\mu\nu}(x)=
\partial_{\nu}\partial_{[\alpha}\hat{h}_{\beta]\mu}
 -\partial_{\mu}\partial_{[\alpha}\hat{h}_{\beta]\nu}.
\end{equation}
Here the convention for antisymmetrization as found in Ref.~\cite{MTW} is
used, i.e.\ $\partial_{\nu}\partial_{[\alpha}h_{\beta]\mu}=
\frac{1}{2}(\partial_{\nu}\partial_{\alpha}h_{\beta\mu} -
\partial_{\nu}\partial_{\beta}h_{\alpha\mu})$.  Henceforth $h_{\mu\nu}$
and $R_{\alpha\beta\mu\nu}$ are understood to be operators, and the
hat may be suppressed without confusion. The expectation value of
$R_{\alpha\beta\mu\nu}$ is
\begin{equation}
 \langle R_{\alpha\beta\mu\nu}(x)\rangle =
 \langle\partial_{\nu}\partial_{[\alpha}h_{\beta]\mu}\rangle -
 \langle\partial_{\mu}\partial_{[\alpha}h_{\beta]\nu}\rangle \,.
\end{equation}
It is convenient to define
\begin{eqnarray}
 K_{\alpha\beta\mu\nu\,\gamma\delta\rho\sigma}(x,x') &=& 
 \langle\partial_{\nu}\partial_{\alpha}h_{\beta\mu}(x)
 \partial_{\sigma}'\partial_{\gamma}'h_{\delta\rho}
 (x')\rangle -
 \langle\partial_{\nu}\partial_{\alpha}h_{\beta\mu}(x)\rangle
 \langle\partial_{\sigma}'\partial_{\gamma}'
 h_{\delta\rho}(x')\rangle \nonumber \\
&=& \partial_{\nu}\partial_{\alpha}\partial_{\sigma}'
 \partial_{\gamma}'
 \left(\langle h_{\beta\mu}(x)h_{\delta\rho}(x')\rangle
 - \langle h_{\beta\mu}(x)\rangle
 \langle h_{\delta\rho}(x')\rangle\right) \, ,
\end{eqnarray}
where $\partial'$ denotes differentiation with respect to
$x^{\prime}$. The Riemann tensor correlation function may be expressed as
\begin{equation} \label{CorFun}
 C_{\alpha\beta\mu\nu\,\gamma\delta\rho\sigma}(x,x') =
 4 K_{[\alpha\beta][\mu\nu]\,[\gamma\delta][\rho\sigma]}(x,x')\,.
\end{equation}

\section{Gravitons in a Squeezed State}
\label{sec;squeeze}
\subsection{Single Mode Squeezed State}
 
Gravitons in a squeezed state are the natural result of quantum
graviton creation in a background gravitational field, such as in
an expanding universe~\cite{GS90}. Here we suppose that the region between 
a source and a detector is filled with gravitons in a squeezed state, which
produce spacetime geometry fluctuations.
 A brief summary of the properties of squeezed
states needed for the following calculations may be found in the
appendix.    
To calculate $C_{\alpha\beta\mu\nu\,\gamma\delta\rho\sigma}(x,x')$, 
we must calculate $\langle h_{\beta\mu}(x)h_{\delta\rho}(x')\rangle$ 
and $\langle h_{\beta\mu}(x)\rangle$. 
We are interested in the change in quantum 
fluctuations between the vacuum state and a squeezed state, so we
may take the correlation function to be normal-ordered.
  We choose to evaluate the expectation
value with respect to a gravity wave in a general single mode
squeezed state $\vert\alpha,\zeta\rangle$ (see the appendix), 
where $\alpha$ and $\zeta$ are the
displacement and squeeze parameters, respectively, and the 
only mode excited  has
a specific wave vector $\ell^\mu$, frequency $\omega_g=\ell^0$ and 
polarization $p$. We may
now proceed to calculate $\langle\zeta,\alpha\vert
h_{\mu\nu}\vert\alpha,\zeta\rangle$, $\langle\zeta,\alpha\vert
h_{\mu\nu}(x)h_{\alpha\beta} (x')\vert\alpha,\zeta\rangle$ and then find
$:K_{\alpha\beta\mu\nu\,\gamma\delta\rho\sigma}(x,x'):$. The result
is:
\begin{equation}
 :K_{\alpha\beta\mu\nu\,\gamma\delta\rho\sigma}(x,x'):
 =\ell_{\alpha}\ell_{\nu}\ell_{\gamma}\ell_{\sigma}A_{\beta\mu}
 A_{\delta\rho}f(x,x')\,,
\end{equation}
where
\begin{equation}
 f(x,x') =
 [\cosh(2r)-1] \cos[\ell_{\tau}(x^{\tau}-x^{\prime\tau})]-
 \sinh(2r)\cos[\ell_{\tau}(x^{\tau}+x^{\prime\tau})+\theta].
\end{equation}
The parameters $r$ and $\theta$ are defined such that
$\zeta=re^{i\theta}$ (see the appendix). Using
Eq.~(\ref{CorFun}), the normal ordered Riemann tensor correlation 
function can be expressed as:
\begin{equation}
 :C_{\alpha\beta\mu\nu\,\gamma\delta\rho\sigma}(x,x'): = 4\;
 (\ell_{[\alpha}A_{\beta][\mu}\ell_{\nu]})(\ell_{[\gamma}A_{\delta ][
 \rho}\ell_{\sigma]})f(x,x').
\end{equation}
Note that the correlation function depends only upon the squeeze parameter
$\zeta$, and not upon the coherent state parameter $\alpha$.

\begin{figure}
  \centering
  \scalebox{.5}{\includegraphics{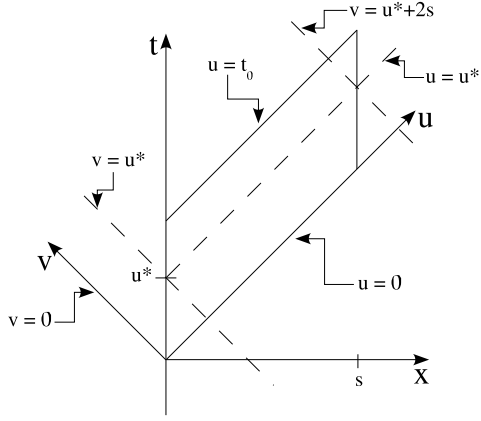}}\\
  \caption{The spacetime region between the source and detector in 
  null coordinates.
  For each fixed $u = u^*$ between $u = 0$ and $u = t_0$, one integrates
  along the line $u = u^*$ from $v = u^*$ to $v = u^*+2s$.}
  \label{spacetime2}
\end{figure}

We now use this normal ordered version of the Riemann tensor correlation
function in Eqs.~(\ref{eq:delta_xi}) and (\ref{eq:delta_theta}). 
When performing the
$da$ and $da'$ spacetime integrations on $f(x,x')$, it is convenient to
use null coordinates, $u=t-x$, $v=t+x$ (see Fig.~\ref{spacetime2}).
The result is
\begin{eqnarray} 
  F(\omega_g,s,t_0) &=& \int da \int da' \, f(x,x') =  \int_0^{t_{0}}du
  \int_0^{t_{0}} du' \int_u^{u+2s} dv \int_{u'}^{u'+2s} dv' \,
  f(x,x')     \nonumber \\
 &=& \frac{16\{1-\cos[s(\ell_x-\omega_g)] \} \,[1-\cos(\omega_g t_0)]}
       {\omega_g^2(\ell_x-\omega_g)^2} \nonumber \\
 &\times& [\cosh(2r)-\sinh(2r)\cos(\theta+s \ell_x-\omega_g(t_{0}+s))-1]\,.
    \label{F}
\end{eqnarray}
Here $s$ is the spatial separation of source and detector, and $t_0$
is the observation time. It should be noted that, since we integrate
over a spacetime slice of constant y and z,  we may 
set $y=y'=z=z'=0$. Finally, from Eq.~(\ref{eq:delta_xi}), the
redshift fluctuations become:
\begin{equation}
  \delta\xi^2 =
  4\, (\ell_{[\alpha} A_{\beta ][
  \mu}\ell_{\nu]})(\ell_{[\gamma}A_{\delta ] [
  \rho}\ell_{\sigma]})t^{\alpha}k^{\beta}t^{\mu}k^{\nu}
  t^{\gamma}k^{\delta}t^{\rho}k^{\sigma}F(\omega_g,s,t_0).
\end{equation}
While from Eq.~(\ref{eq:delta_theta}), the angular fluctuations are given by
\begin{equation}
  \delta\Theta^2 =
  4\, (\ell_{[\alpha} A_{\beta ][
  \mu}\ell_{\nu]})(\ell_{[\gamma}A_{\delta ][
  \rho}\ell_{\sigma]})s^{\alpha}k^{\beta}t^{\mu}k^{\nu}s^{\gamma}k^{\delta}
  t^{\rho}k^{\sigma}F(\omega_g,s,t_0).
\end{equation}
Interestingly, since $F(\omega_g,s,t_0)$ is independent of the
displacement parameter, $\alpha$, so are $\delta\xi^2$ and
$\delta\Theta^2$. Therefore, the fluctuations depend only on the
squeezing parameter, $\zeta$, and we can immediately say that a coherent
state (classical wave) for which $\zeta=0$ induces no fluctuations. For
the following calculations, we assume the passing gravity waves are
in the Transverse Tracefree~(TT) Gauge. However, note that since our
results derive from the Riemann tensor, the equations are gauge
invariant, and we use the TT gauge only for calculational convenience.

To begin, note that if the gravity waves and photons travel in
parallel, then one finds $(\ell_{[\alpha} A_{\beta ][ \mu}\ell_{\nu]})
t^{\alpha}k^{\beta}t^{\mu}k^{\nu} = 0$.  This indicates that there
is no induced redshift fluctuation nor angular blurring due to a
squeezed state gravity wave traveling with the photons.  This is as
one might expect, since a gravity wave in the TT gauge has only
non-zero transverse components.  We may subsequently restrict our
attention to transversely propagating gravity waves and will assume
the gravity waves to propagate with $\ell^{\mu}=\omega_g(1,0,0,1)$
while the photons continue to have wave vector $k^{\mu}=(1,1,0,0)$.
We define
\begin{equation}
 \mathfrak{F}(\omega_g,s,t_0)= [1-\cos(\omega_g s)][1-\cos(\omega_g t_0)]
 [\cosh(2r)-\sinh(2r)\cos(\theta-\omega_g(t_{0}+s))-1]\, ,
\end{equation}
so that for a gravitational wave propagating in the z direction, for which
$\ell_x=0$,
\begin{equation}
 F(\omega_g,s,t_0) = \frac{16}{{\omega_g}^4}\mathfrak{F}(\omega_g,s,t_0) \,.
\end{equation}

\subsubsection{Redshift Fluctuations} \label{rfcase}

With transversely propagating gravity waves,
\begin{equation} (\ell_{[\alpha} A_{\beta ][
  \mu} \ell_{\nu]})t^{\alpha}k^{\beta}t^{\mu}k^{\nu} = 
-\frac{1}{4}\omega_g^2 A_{+}\, , 
\end{equation}
and therefore
\begin{equation} \label{rfcase2eq}
 \delta\xi^2 =  4\, A_{+}^2 \mathfrak{F}(\omega_g,s,t_0).
\end{equation}
There exists a non-zero effect which in this case depends only on
the $(+)$ polarization. For a gravity wave propagating in the $z$-direction,
$A_+ = A_{xx} =-A_{yy}$.

\subsubsection{Angular Blurring} \label{abcase}
\paragraph{Case 1} 
Let the photons
propagate in the x direction while the gravity waves propagate in
the z direction and probe the y component of angular blurring,
$s^{\mu}=(0,0,1,0)$. We find
\begin{equation} (\ell_{[\alpha} A_{\beta ][
  \mu}\ell_{\nu]})s^{\alpha}k^{\beta}t^{\mu}k^{\nu} = 
\frac{1}{4}\omega_g^2 A_{\times}
\end{equation}
and
\begin{equation} \label{abcase2eq}
 \delta\Theta_y^2 = 4\, A_{\times}^2\mathfrak{F}(\omega_g,s,t_0).
\end{equation}
Note that here the y component of blurring depends only on the
$(\times)$ polarization of the gravity waves. 
Here $A_{\times} = A_{xy}=A_{yx}$.

\paragraph{Case 2} 
To probe the z component of blurring, set
$s^{\mu}=(0,0,0,1)$. The result is
\begin{equation} \label{abcase3eq}
 \delta\Theta_z^2 = 4\, A_{+}^2\mathfrak{F}(\omega_g,s,t_0)\, ,
\end{equation}
which is the same as Eq.~(\ref{rfcase2eq}) for redshift fluctuations,
and only depends on the $(+)$ polarization of the gravity waves.

\subsection{Classical Time Dependent Variation}

As noted above, the fluctuations in redshift and angle depend upon
the degree of squeezing, measured by the parameter $\zeta$.  In this
subsection, we will examine the expectation value of the change in
fractional redshift, $\langle\Delta\xi\rangle$ and in angle,
$\langle\Delta\Theta\rangle$. These quantities will depend only
upon the displacement parameter $\alpha$, and hence would be the same
in the coherent state $|\alpha,0 \rangle$ as they are in the squeezed
state  $|\alpha,\zeta \rangle$. For this reason, we regard these
quantities as giving the classical time-dependence.  Alternative approaches 
to the classical time-dependence of these quantities may be found in 
Refs.~\cite{Weber, Zipoy, Kaufmann}

Here it is possible to
calculate $\langle\Delta\xi\rangle$ as a single integration of the
Riemann tensor over the spacetime region $da$ via 
Eq.~(\ref{RateOfChange}). However, for comparison purposes, we calculate
$\langle\Delta\xi\rangle^2$ directly. In this case,
\begin{equation}
 \langle\Delta\xi\rangle^2 = \int da \int da'
 \,\langle\hat{R}_{\alpha\beta\mu\nu}(x)\rangle
 \langle\hat{R}_{\gamma\delta\rho\sigma}(x')\rangle
 t^{\alpha}k^{\beta}t^{\mu}k^{\nu}t^{\gamma}k^{\delta}t^{\rho}k^{\sigma}\, ,
\end{equation}
where now
\begin{equation}
 \langle\hat{R}_{\alpha\beta\mu\nu}(x)\rangle
 \langle\hat{R}_{\gamma\delta\rho\sigma}(x')\rangle = 4 \sum_{l}
 (\ell_{[\alpha} A_{\beta][\mu}\ell_{\nu]})(\ell_{[\gamma}
 A_{\delta][\rho}\ell_{\sigma]}) f'(x,x')
\end{equation}
with
\begin{equation}
 f'(x,x') =
 \alpha^2 e^{i\ell_{\tau} (x^{\tau}+x^{\prime\tau})} +
 (\alpha^{\ast})^2 e^{-i\ell_{\tau} (x^{\tau}+x^{\prime\tau})} +
 2\vert\alpha\vert^2\cos[\ell_{\tau} (x^{\tau}-x^{\prime\tau})].
\end{equation}
Integration of $f'(x,x')$ over the spacetime region of 
Fig.~\ref{spacetime} yields
\begin{eqnarray}
& & F'(\omega_g,s,t_0) = \int da \int da' \, f'(x,x') = \nonumber \\
& & \frac{16\{1-\cos[s(\ell_x-\omega_g )]\} [1-\cos(\omega_g t_0)]}
 {\omega_g^2(\ell_x-\omega_g)^2}  
 \left(\alpha e^{\frac{i}{2}[\ell_x s-\omega_g(s+t_0)]}
 + \alpha^{\ast} e^{-\frac{i}{2}[\ell_x s-\omega_g(s+t_0)]}\right)^2.
\end{eqnarray}
Thus the classical time variation of the redshift is characterized by:
\begin{equation}\label{classical}
 \langle\Delta\xi\rangle^2 =
 4\, F'(\omega_g,s,t_0)
 (\ell_{[\alpha} A_{\beta][\mu}\ell_{\nu]})(\ell_{[\gamma}
 A_{\delta][\rho}\ell_{\sigma]})
 t^{\alpha}k^{\beta}t^{\mu}k^{\nu}t^{\gamma}k^{\delta}t^{\rho}k^{\sigma}.
\end{equation}
For the case of redshift variations with photons and a gravity wave
propagating perpendicularly (Sect.~\ref{rfcase}),
\begin{equation} \label{rfclassical}
 \langle\Delta\xi\rangle^2 =
  4 \,A_{+}^2 \mathfrak{F'}(\omega_g,s,t_0).
\end{equation}
Here $\mathfrak{F'}(\omega_g,s,t_0)$ is defined in a similar way as
$\mathfrak{F}(\omega_g,s,t_0)$ when $\ell_x=0$,
\begin{equation}
 \mathfrak{F'}(\omega_g,s,t_0) =  [1-\cos(\omega_g s)][1-\cos(\omega_g t_0)]
 \left(\alpha e^{-\frac{i\omega_g}{2}(s+t_0)}
 + \alpha^{\ast} e^{\frac{i\omega_g}{2}(s+t_0))}\right)^2.
\end{equation}
The classical time variation of angular position yields similar
results. Particularly, for perpendicularly propagating photons and
gravity waves and probing the y component of blurring (case 1 of
Sect.~\ref{abcase}), one finds
\begin{equation} \label{abclassical1}
 \langle\Delta\Theta_y\rangle^2 = 4\,A_{\times}^2 \mathfrak{F}'(\omega_g,s,t_0).
\end{equation}
While probing the z component of blurring (case 2 of Sect.~\ref{abcase}),
the result is
\begin{equation} \label{abclassical}
 \langle\Delta\Theta_z\rangle^2 = 
 4 \,A_{+}^2 \mathfrak{F}'(\omega_g,s,t_0).
\end{equation}
Note that the function $F'(\omega_g,s,t_0)$, and hence 
$\mathfrak{F'}(\omega_g,s,t_0)$, depends on the
displacement parameter $\alpha$, but not on the squeeze parameter,
$r$.  Therefore the same can be said for results
Eqs.~(\ref{rfclassical}), (\ref{abclassical1}), and (\ref{abclassical}).
In particular $\mathfrak{F'}(\omega_g,s,t_0)=0$ for $\alpha=0$. This means a
coherent state (classical wave), for which $r=0, \alpha\neq 0$,
exhibits regular time variations but no fluctuations.  Indeed,
recall from Eq.~(\ref{F}) that $\delta\xi^2=\delta\Theta^2=0$
for $r=0$.  We will exploit this fact shortly by considering a state
for which $\alpha = 0$.

In a time-averaged measurement, the classical time variation will
produce line broadening and angular blurring, such as do the quantum
spacetime fluctuations from a squeezed vacuum state. The two effects can, 
however, be distinguished in principle. If one were to make repeated
measurements at the same point in the cycle of a gravity wave, the
effects of classical time dependence would disappear, but those of
quantum fluctuations would remain.

\subsection{Stress Tensor for Squeezed State Gravity Waves} 

We
would like to obtain an order of magnitude estimate for
$\langle\left(\Delta\xi\right)^2\rangle$ and
$\langle(\Delta\Theta)^2\rangle$. However, the gravitational wave
amplitude $A_{\mu\nu}$ is not directly measurable.  We therefore
would like to express the previous results in terms of the energy
density of the waves.  While the stress-energy tensor is not well
defined for a gravitational wave, one can define an effective stress
tensor in the linearized theory (see, e.g., Ref.~\cite{MTW}).
Classically, this effective stress tensor is defined as
\begin{equation}
 T_{\mu\nu}^{GW} = \frac{1}{32 \pi}\, \langle
 h^{TT}_{\alpha\beta,\mu}h^{TT}_{\alpha\beta,\nu} \rangle\, .
\end{equation}
Here the brackets denote a spatial average over several
wavelengths and the TT superscript indicates we are working in the
Transverse Tracefree gauge. We use this expression with the
expectation value of a gravity wave in a squeezed state,  and
perform a spatial average on the quantity
\begin{equation}
  \langle\zeta,\alpha\vert :\hat{h}^{TT}_{\alpha\beta,\mu}
  \hat{h}^{TT}_{\alpha\beta,\nu}:
  \vert\alpha,\zeta\rangle
\end{equation}
This operation results in an effective stress tensor of the form
\begin{equation}
 T_{\mu\nu} = \frac{1}{32 \pi}\, \sum_{\alpha,\beta}A_{\alpha\beta}^2
 k_{\mu}k_\nu [2\vert\alpha\vert^2+\cosh(2r)]
\end{equation}
while
\begin{equation}\label{T}
 :T_{\mu\nu}: \, = \frac{1}{32 \pi}\,\sum_{\alpha,\beta}A_{\alpha\beta}^2
 k_{\mu}k_\nu [2\vert\alpha\vert^2+\cosh(2r)-1].
\end{equation}
The vacuum energy term, $T_{00}^{vac} = T_{00}-:T_{00}:$ is then
\begin{equation}
 T_{00}^{vac} =
 \frac{1}{32 \pi}\,\sum_{\alpha,\beta}A_{\alpha\beta}^2\omega_g^2\,.
\end{equation}
Examining the contribution from the $(+)$ polarization, where
$A_+ = A_{xx} = -A_{yy}$ and $A_\times = A_{xy} = A_{yx} = 0$, one finds
\begin{equation}
 T_{00}^{vac} = \frac{1}{16 \pi}\, A_{+}^2\omega_g^2.
\end{equation}
However, since the quantum vacuum energy is also
$\frac{1}{2}\omega_g$, we can use the vacuum mode to fix the
normalization.
\begin{equation}
 VT_{00}^{vac}= \frac{1}{16 \pi}\, A_{+}^2\omega_g^2V=\frac{1}{2}\omega_g
\end{equation}
for each $\bm{\ell}$, where $V$ is the quantization volume,  which leads to
\begin{equation}\label{Amplitude}
 A_{+}= \sqrt{\frac{8 \pi}{\omega_gV}}.
\end{equation}
For the $(\times)$ polarization, one also finds
\begin{equation}
 A_{\times}= \sqrt{\frac{8 \pi}{\omega_gV}}.
\end{equation}
The previous results, Eqs.~(\ref{rfcase2eq}),
(\ref{abcase2eq}), and (\ref{abcase3eq}) become
\begin{equation}
 \delta\xi^2 = \delta\Theta_y^2 = \delta\Theta_z^2 =
 \frac{32 \pi}{\omega_gV} \, \mathfrak{F}(\omega_g,s,t_0)\,.
\end{equation}
So far, we have considered a single mode. However, we may also have a 
situation in which many modes are excited. In this case, we would insert a
sum on modes on the right hand side of the above equation.
If the density of states is large, we let
\begin{equation}
 \sum_{\ell} \rightarrow \frac{V}{(2\pi)^3}\int{d^3 \bm{\ell}}. 
\end{equation}
Suppose the distribution of gravity waves in 
$\bm{\ell}$-space is narrowly
peaked about some $\omega_g$ with characteristic width
$\Delta\omega_g$. If $\Delta\omega_g$ is small, then the integrand
is essentially constant over the region of integration. The result
of the integration is just the integrand multiplied by a volume,
$(\Delta \ell_x)(\Delta \ell_y)(\Delta
 \ell_z)$, in $\bm{\ell}$-space, with 
$\Delta \ell_i$ a bandwidth in  the $i$-direction of
$\bm{\ell}$-space. Specifically,
\begin{equation} \label{Eint}
 \delta\xi^2 = \frac{\omega_g^3}{4 \pi^2}\,  
(\Delta \ell_x)(\Delta \ell_y)(\Delta \ell_z)\, F(\omega_g,s,t_0).
\end{equation}

\subsection{Estimate of $\delta\xi^2$ and $\delta\Theta^2$} 

Recall that from Eqs.~(\ref{rfclassical}),
(\ref{abclassical1}), and (\ref{abclassical}), the classical time
variation depends on the displacement parameter, $\alpha$, but not
on the squeeze parameter $\zeta$.  For the purpose of obtaining order of
magnitude estimates of $\delta\xi^2$ and $\delta\Theta^2$, suppose
$\alpha = 0$ and $r \gg 1$; further assume the $+$ polarization, so
that $A_{\times}=0$. Since for $\alpha = 0$, 
$\mathfrak{F'}(\omega_g,s,t_0) = 0$,
it is clear that $\langle\Delta\xi\rangle^2 = 0$ and
$\langle\Delta\Theta\rangle^2 = 0$, from Eqs.~(\ref{rfclassical}) and
(\ref{abclassical}), respectively. Therefore, for the case
$\alpha=0$ one finds $\delta\xi^2 = \langle(\Delta\xi)^2\rangle$ and
$\delta\Theta^2 = \langle(\Delta\Theta)^2\rangle$.

Using Eqs.~(\ref{T}) and (\ref{Amplitude}), and integrating over 
a sharply peaked distribution
function in $\bm{\ell}$-space, the energy density for large $r$ becomes
\begin{equation}
 :T_{00}: \, \approx 
\frac{\omega_g e^{2r}}{4(2\pi)^3}\Delta \ell_x \Delta \ell_y
 \Delta \ell_z.
\end{equation}
  From Eqs.~(\ref{F}) and (\ref{Eint}), we have for large $r$,
\begin{equation}
 \langle(\Delta\xi)^2\rangle \approx
 \frac{2 \,e^{2r}}{\pi^2 \omega_g}\Delta \ell_x \Delta \ell_y \Delta \ell_z.
\end{equation}
Thus
\begin{equation}
 \langle(\Delta\xi)^2\rangle \approx
 64 \pi \frac{:T_{00}:}{\omega_g^2}  =
 64 \pi \frac{:T_{00}:}{\omega_g^2}\ell_{\rm Pl}^2\, ,
\end{equation}
where $\ell_{\rm Pl}$ is the Planck length.
With $\omega_g = 2 \pi/ \lambda_g$, we have
\begin{equation}
 \langle(\Delta\xi)^2\rangle \approx
\frac{16}{\pi} \ell_{\rm Pl}^2\lambda_g^2 :T_{00}: 
\quad \mbox{or} \quad 
 (\Delta\xi)_{rms}\approx
 \ell_{\rm Pl} \lambda_g \sqrt{:T_{00}:} \, .
\end{equation}
Suppose for example a gravity wave with $\lambda_g = 1\,\mbox{ly}
=10^{18} \mbox{cm}$ and the closure energy density $:T_{00}:\,=10^{8}
\mbox{cm}^{-4}$.  Then with $\ell_{\rm Pl}=10^{-33}\mbox{cm}$,
\begin{equation} \label{magnitude}
 (\Delta\xi)_{rms} \approx 10^{-33}
 10^{18} \sqrt{10^8} = 10^{-11}\,.
\end{equation}
By comparing Eqs.~(\ref{rfcase2eq}) and (\ref{abcase3eq}), the
order of magnitude of $(\Delta\Theta)_{rms}$ will be the
same as the result for $(\Delta\xi)_{rms}$.

In principle, the effect can be made as large as desired by
increasing the squeezing parameter $r$, which increases the energy
density of the wave.  However, as shown from the estimate
given, this is a very small effect and is likely to be unobservable
in the present day universe. This example serves as a useful model for
spacetime geometry fluctuations, which will have large effects in the early
universe and in the vicinity of black holes.

\section{Thermal Bath of Gravitons}
\label{sec:thermal}

In the previous section, the fluctuation effects of a gravitational
wave in a single mode squeezed state were examined.  Another useful
example is a thermal bath of gravitons, such as might be created by
an evaporating black hole.  We now consider such a
thermal bath as the source of spacetime fluctuations.
In particular, suppose the spacetime geometry fluctuates in such a
way that $\langle\Gamma_{\alpha\beta}^{\mu}\rangle = \langle
R^{\mu}_{\phantom{\mu}\alpha\nu\beta}\rangle = 0$, but $\langle
R^{\mu}_{\phantom{\mu}\alpha\nu\beta}R^{\gamma}_{\phantom{\gamma}
\delta\rho\sigma}\rangle_{\beta} \neq 0$.  In effect, we are
ignoring the average spacetime curvature due to the bath of
gravitons. In this case the Riemann tensor correlation function is
\begin{equation}
 C_{\alpha\beta\mu\nu\,\gamma\delta\rho\sigma}(x,x') = 
\langle R_{\alpha\beta\mu\nu}(x)
 R_{\gamma\delta\rho\sigma}(x')\rangle_{\beta}\, ,
\end{equation}
and therefore
\begin{equation} \label{thermalRF1}
 \delta\xi^2 = \langle(\Delta\xi)^2\rangle = \int da \int da' \, \langle
 R_{\alpha\beta\mu\nu}(x)
 R_{\gamma\delta\rho\sigma}(x')\rangle_{\beta} \,
 t^{\alpha}k^{\beta}t^{\mu}k^{\nu}t^{\gamma}k^{\delta}t^{\rho}k^{\sigma}
\end{equation}
and
\begin{equation} \label{thermalAB1}
 \delta\Theta^2 = \langle(\Delta\Theta)^2\rangle = \int da \int da' \,
 \langle R_{\alpha\beta\mu\nu}(x)
 R_{\gamma\delta\rho\sigma}(x')\rangle_{\beta} \,
 s^{\alpha}k^{\beta}t^{\mu}k^{\nu}s^{\gamma}k^{\delta}t^{\rho}k^{\sigma}\, ,
\end{equation}
where
$\langle R_{\alpha\beta\mu\nu}(x)
R_{\gamma\delta\rho\sigma}(x')\rangle_{\beta}$ is the thermal normal-ordered
Riemann tensor two point function.

\subsection{Redshift Fluctuations}

In the average rest frame of the source and detector, let $v^{\mu} =
t^{\mu} = (1,0,0,0)$ and $k^{\mu} = (1,1,0,0)$ as
previously.  With this and the symmetry and cyclic properties of the
Riemann tensor, Eq.~(\ref{thermalRF1}) reduces to
\begin{equation}
 \langle(\Delta\xi)^2\rangle = \int da \int da' \,
 \langle R_{0101}(x)R_{0101}(x')\rangle_{\beta}.
\end{equation}
We construct the thermal Riemann tensor two point function via the
Matsubara method as an infinite image sum in imaginary time of the 
vacuum two point function.  We proceed for the moment in the
Feynman gauge, but since we are strictly dealing with functions of
the Riemann tensor, the result will be gauge independent. From
Eq.~(\ref{MetricDerivatives}) and the symmetry properties of
the metric tensor, the Riemann tensor vacuum two-point function can
be written as
\begin{multline}
  \langle R_{0101}(x)R_{0101}(x')\rangle \\ \shoveleft{=\frac{1}{4}\langle(
  2h_{01,01} - h_{00,11} - h_{11,00})(x)(2h_{01,01} - h_{00,11} - h_
  {11,00})(x')\rangle} \\ = \frac{1}{4}[4G_{0101,0101} -
  2G_{0100,0111}-2G_{0111,0100} -2G_{0001,1101} + G_{0000,1111}
  \\ + G_{0011,1100} -2G_{1101,0001} + G_{1100,0011} +
  G_{1111,0000}].
\end{multline}
Here, $G_{\alpha\beta\mu\nu}$ is the metric two point function.  The
first two indices on $G_{\alpha\beta\mu\nu}$ refer to point $x$
while the second two refer to point $x'$ and similarly for the
derivative indices.  In the Feynman gauge
\begin{equation}
 G_{\alpha\beta\mu\nu} = (\eta_{\alpha\mu}\eta_{\beta\nu}+
 \eta_{\alpha\nu}\eta_{\beta\mu} -
 \eta_{\alpha\beta}\eta_{\mu\nu})D
\end{equation}
where $D$ is the vacuum scalar two point function
\begin{equation}
 D = \frac{1}{4\pi^2}\frac{1}{(\Delta \vec{x})^2 - (\Delta t)^2}\,.
\end{equation}
 One finds that the only nonzero components of 
$G_{\alpha\beta\mu\nu}$ are $G_{0101}, G_{0000}, G_{0011},
G_{1100}$, and $G_{1111}$ and thus
\begin{equation}
 \langle R_{0101}(x)R_{0101}(x')\rangle = \\
 \frac{1}{4}[-4D,_{txt'x'} + D,_{xxx'x'} + D,_{xxt't'} +
 D,_{ttx'x'} +D,_{ttt't'}].
\end{equation}
One can easily see from the function $D$ that $D,_{txt'x'} =
D,_{xxt't'} = D,_{ttx'x'} = \partial_t^2\partial_x^2D$ and therefore
\begin{equation}
 \langle R_{0101}(x)R_{0101}(x')\rangle =
 \frac{1}{4}(\partial_t^4 - 2\partial_t^2\partial_x^2 +
 \partial_x^4)D.
\end{equation}
It is now clear that the thermal Riemann tensor two point function
may be constructed from the vacuum two point function by making the
replacement $D\rightarrow D_{\beta}$ whence
\begin{equation} \label{ThermalRFCorr}
 \langle R_{0101}(x)R_{0101}(x')\rangle_{\beta} =
 \frac{1}{4}(\partial_t^4 - 2\partial_t^2\partial_x^2 +
 \partial_x^4)D_{\beta}.
\end{equation}
As mentioned earlier, $D_{\beta}$ is constructed via the Matsubara
method. First make $D$ periodic in imaginary time and then take an
infinite image sum so that
\begin{equation}
 D_{\beta} = 
\frac{1}{4\pi^2}\sideset{}{'}{\sum}_{n=-\infty}^{\infty}
\frac{1}{(\Delta \vec{x})^2 -
 (t-t'+in\beta)^2}.
\end{equation}
The prime on the summation indicates that we leave out the $n=0$
term, which corresponds to the vacuum term. As a result, 
Eq.~(\ref{ThermalRFCorr})
gives a normal-ordered two-point function. We can set $y-y' = z-z' = 0$,
and  use null coordinates to write
\begin{equation}
 D_{\beta} = \frac{1}{4\pi^2}\sideset{}{'}{\sum}_{n=-\infty}^{\infty}\frac{1}{n^2\beta^2 - \Delta v
 \Delta u - in\beta(\Delta v + \Delta u)}.
\end{equation}
This, with the null coordinate version of Eq.~(\ref{ThermalRFCorr}), gives
\begin{equation}
 \langle(\Delta\xi)^2\rangle = 4 \int da \int da' \,
 \partial_u\partial_{u'}\partial_v\partial_{v'}D_{\beta}.
\end{equation}
The integral can be evaluated in the following way.  First note
that we are interested in the real part of $D_{\beta}$, which is
even in $n$, so that we may make the replacement
\[\sideset{}{'}{\sum}_{n=-\infty}^{\infty} \rightarrow
2\sum_{n=1}^{\infty}\] and let
\begin{equation}
 Re (D_{\beta}) = \frac{1}{2\pi^2}\sum_{n=1}^{\infty} G
\end{equation}
where
\begin{equation}
 G = Re \left[\frac{1}{n^2\beta^2 - \Delta v
 \Delta u - in\beta(\Delta v + \Delta u)}\right].
\end{equation}
In null coordinates,
\begin{equation}
 \langle(\Delta\xi)^2\rangle = 
\frac{2}{\pi^2}\sum_{n=1}^{\infty}\int_0^{t_{0}}du
 \int_0^{t_{0}} du' \int_u^{u+2s} dv \int_{u'}^{u'+2s} dv' \,
 \partial_v\partial_{v'} \partial_u\partial_{u'}G.
\end{equation}
Performing the integrations on $v$ and $v'$ yields
\begin{equation}
 \langle(\Delta\xi)^2\rangle = 
\frac{2}{\pi^2}\sum_{n=1}^{\infty}\int_0^{t_{0}}du
 \int_0^{t_{0}} du' \, I(\Delta u)\, ,
\end{equation}
where we make use of the fact that $\partial_u\partial_{u'}G =
-\partial_u^2G$ in writing
\begin{equation}
 I(\Delta u) = (\partial_u^2 G)\Bigr|_{\Delta v = \Delta u+2s} +
 (\partial_u^2 G)\Bigr|_{\Delta v = \Delta u-2s} - 2(\partial_u^2 G)\Bigr|_{\Delta v = \Delta u}.
\end{equation}
The function $I(\Delta u)$ is a function of $\Delta u$ only, for
which
\begin{equation}
 \int_0^{t_0} du \int_0^{t_0} du' \, I(\Delta u) = \int_0^{t_0} d(\Delta
 u)(t_0-\Delta u)I(\Delta u)+\int_{-t_0}^0 d(\Delta
 u)(t_0+\Delta u)I(\Delta u).
\end{equation}
This now leads to the expression
\begin{equation} \label{gravrf}
 \langle(\Delta\xi)^2\rangle =
 \frac{2}{\pi^2}\sum_{n=1}^{\infty}\int_0^{t_0} d(\Delta
 u)(t_0-\Delta u)I(\Delta u)+\int_{-t_0}^0 d(\Delta
 u)(t_0+\Delta u)I(\Delta u).
\end{equation}
Equation~(\ref{gravrf}) can be evaluated using a program such as
Maple.  The complete result is a rather lengthy expression and
no insight is gained by writing it out.  However, in the limit where
$s \gg t_0$ and $s \gg\beta$, i.e., the observationally reasonable limit where
the distance between source and detector is much larger than both
the observation time and the thermal wavelength, the expression
reduces to
\begin{equation}
 \langle(\Delta\xi)^2\rangle\sim
 \frac{4}{9\pi^2 a^2\beta^2}[3\pi^2 a^2
 \mbox{csch}^2(\pi a)+
 \pi^2 a^2 - 3],
\end{equation}
with $a={t_0}/{\beta}$. In the limit where the observation time
is short compared to the correlation time $\beta$, i.e.\ $t_0 \ll
\beta$, we find
\begin{equation}
 \langle(\Delta\xi)^2\rangle \sim \frac{4t_0^2\pi^2}{45\beta^4} =
 \frac{4t_0^2\pi^2\ell_{\rm Pl}^2}{45\beta^4}
\end{equation}
In this case, the rms line width grows linearly with observation
time,
\begin{equation}
 (\Delta\xi)_{rms} = \frac{2\pi\ell_{\rm Pl} t_0}{\beta^2\sqrt{45}}.
\end{equation}
While on the other hand if $t_0 \gg \beta$,
\begin{equation}
 \langle(\Delta\xi)^2\rangle \sim \frac{4}{9\beta^2} =
 \frac{4\ell_{\rm Pl}^2}{9\beta^2}.
\end{equation}
Here the rms line width approaches a constant,
\begin{equation}
 (\Delta\xi)_{rms} = \frac{2\ell_{\rm Pl}}{3\beta}.
\end{equation}

\subsection{Angular Blurring}

Turning our attention to Eq.~(\ref{thermalAB1}), once again let
$v^{\mu} = t^{\mu} = (1,0,0,0)$ and $k^{\mu} = (1,1,0,0)$;
additionally let $s^{\mu}=(0,0,1,0)$. With this and the symmetry and
cyclic properties of the Riemann tensor, Eq. (\ref{thermalAB1})
reduces to
\begin{equation}
 \langle(\Delta\Theta)^2\rangle = \int da \int da' \,
 \langle
 (R_{2101}(x)-R_{2010}(x))(R_{2101}(x')-R_{2010}(x'))\rangle_{\beta}.
\end{equation}
Proceeding in the Feynman Gauge, the calculations follow those for
redshift fluctuations.  It is straightforward to show that
\begin{multline}
 \langle(R_{yxtx}(x)-R_{ytxt}(x))(R_{yxtx}(x')-R_{ytxt}(x'))\rangle
 \\ = \frac{1}{4}(\partial_t^4 + 2\partial_t^3\partial_x -
 2\partial_t\partial_x^3 - \partial_x^4)D = -4\partial_u
 \partial_v^2 \partial_{v'}D\,.
\end{multline}
In null coordinates, the equation for angular blurring becomes
\begin{equation}
 \langle(\Delta\Theta)^2\rangle = -4 \int_0^{t_0}du \int_0^{t_{0}}
 du' \int_u^{u+2s} dv \int_{u'}^{u'+2s} dv' \, \partial_u
 \partial_v^2 \partial_{v'}D.
\end{equation}
This integral can be solved via the same method invoked in the
previous subsection when dealing with redshift fluctuations.  While
it is not immediately obvious from this form of the integral, the
limiting results are the same as for line broadening. Namely, in the
case when $s \gg t_0,\beta$ we again have
\begin{equation}
 \langle(\Delta\Theta)^2\rangle\sim
 \frac{4}{9\pi^2 a^2\beta^2}[3\pi^2 a^2
 \mbox{csch}^2(\pi a)+
 \pi^2 a^2 - 3],
\end{equation}
In the limit where $a \rightarrow 0$, or $t_0 \ll \beta$,
\begin{equation}
 (\Delta\Theta)_{rms} \sim \frac{2\pi\ell_{\rm Pl} t_0}{\beta^2\sqrt{45}}\, ,
\end{equation}
while on the other hand when $a \rightarrow \infty$, or $t_0 \gg \beta$,
\begin{equation}
 (\Delta\Theta)_{rms} \sim \frac{2 \ell_{\rm Pl}}{3 \beta} \,.
         \label{eq:theta_therm}
\end{equation}
As emphasized earlier, these results are independent of gauge
choice. As a check, these results were also calculated in the TT
gauge using the thermal two point function and Hadamard function for
the TT gauge derived in Ref.~\cite{Yu}.  

We may compare Eq.~(\ref{eq:theta_therm}) with 
the results of Ref.~\cite{Borgman}, where a heuristic result is found for 
the angular blurring of the image of a source caused by passive fluctuations 
rather than active fluctuations as done here.  The source of fluctuation 
there is taken to be thermal fluctuations in the stress tensor of a scalar 
field.  In the high temperature limit, the passive fluctuation result is 
\begin{equation}
 (\Delta\Theta)_{pass} \sim 
\frac{\ell_{\rm Pl}^2 s^{\frac{3}{2}}}{b \beta^{\frac{5}{2}}} \, ,
\end{equation}
where $b$ is a characteristic width of the bundle of rays.
  The effect is seen to increase with 
source-detector separation, though it should be mentioned that this result 
is valid for the regime $s\alt {\beta^3}/{\ell_{\rm Pl}^2}$.  An important 
point is that the result for passive fluctuations has two powers of 
$\ell_{\rm Pl}$, 
while those for the active fluctuations have only one.  One may therefore 
conclude that the effect tends to be larger for active fluctuations than 
for passive fluctuations.

\section{Summary and Discussion}
\label{sec:final}

In this paper, two effects of spacetime geometry fluctuation
are examined. Expressions are derived for fluctuations of redshift
and angular position of a source as observed by a detector. The
physical manifestation of these effects is found to be a broadening
of observed spectral lines and  angular blurring of the image of
a source. The fluctuation of these observables depends on the
Riemann tensor correlation function, which in turn characterizes
fluctuations in the spacetime geometry. These effects should arise
for both active and passive metric fluctuations. However, in the case
of passive fluctuations, it may be necessary to do some additional
spacetime averaging, as was discussed in Ref.~\cite{Borgman}. The
explicit examples discussed in the present paper concerned active 
fluctuations.

The effects of active spacetime fluctuations are examined by
considering a linearized model of quantum gravity with gravitons
occupying a squeezed vacuum state. In the absence of squeezing there
is no effect, but the effects grow exponentially with the squeezing
parameter and so in principle can be made quite large.  The redshift
and angular position fluctuations are related to the energy density
of the wave, and finally an order of magnitude is estimated for a
gravity wave with large energy density.  The result of this
estimation is quite small with questionable observability in the present day
universe. The
results, however, indicate this is an interesting model for
investigating the properties of  fluctuating spacetime geometries.

Further, a thermal bath of gravitons is considered as the source of
spacetime geometry fluctuation.  An expression for spectral line
broadening is provided for the limit that the source-detector
separation is much larger than both the observation time and the
thermal wavelength.  For observation times short compared to the
thermal wavelength, the rms spectral line width increases linearly
with the observation time.  On the other hand, for observation times
long compared to the thermal wavelength, the rms line width
approaches a constant.  The results for angular blurring mirror the
results for spectral line broadening in the same limit of large
source-detector separation.

\begin{acknowledgments} 
 This work was supported in part by the National
Science Foundation under Grant PHY-0244898.
\end{acknowledgments}

\appendix
\section{Squeezed States} \label{squeezed states}

A squeezed vacuum
state is the natural state for a quantum mechanically created particle
occupying an in-vacuum state represented in an out-Fock space.
Squeezed quantum states are generated via the unitary displacement
and squeeze operators. Here we provide a brief summary of the
relevant ideas and results for squeezed states following primarily
the notation as found in Ref.~\cite{Caves}; see also
Refs.~\cite{Stoler,Stoler2,Yuen}. The displacement operator generates the
set of coherent states and is defined by:
\begin{equation}
 D(\alpha)\equiv \exp(\alpha a^{\dagger} - \alpha^{\ast} a) =
 e^{-\frac{|\alpha|^{2}}{2}}e^{\alpha a^{\dagger}}e^{-\alpha a},
\end{equation}
where $\alpha$ is an arbitrary complex number.  We will here be
primarily interested in generating the coherent state
$\vert\alpha\rangle$ by displacing the vacuum state $\vert 0
\rangle$:
\begin{equation}
\vert\alpha\rangle = D(\alpha)\vert0\rangle\,.
\end{equation}
The displacement operator transforms $a$ and $a^{\dagger}$ as:
\begin{subequations} \label{Disp:gp}
  \begin{gather}
    D^{\dagger}(\alpha)aD(\alpha) = a+\alpha \label{Disp:gp1} \\
    D^{\dagger}(\alpha)a^{\dagger}D(\alpha) = a^{\dagger}+\alpha^{\ast} \label{Disp:gp2}
  \end{gather}
\end{subequations}
The squeeze operator is defined as
\begin{equation}
 S(\zeta) =
 \exp[\frac{1}{2}\zeta^{\ast}a^{2}-\frac{1}{2}\zeta(a^{\dagger})^2],
 \quad \zeta=re^{i\theta},
\end{equation}
where $\zeta$ is an arbitrary complex number.  We use the convention
in \cite{Caves} for the squeeze operator transformations of $a$ and
$a^{\dagger}$:
\begin{subequations} \label{Sq:gp}
  \begin{gather}
    S^{\dagger}(\zeta)aS(\zeta) = 
a\cosh{r}-a^{\dagger}e^{i\theta}\sinh{r} \label{Sq:gp1} \\
    S^{\dagger}(\zeta)a^{\dagger}S(\zeta) = 
a^{\dagger}\cosh{r}-ae^{-i\theta}\sinh{r} \label{Sq:gp2}
  \end{gather}
\end{subequations}
The squeezed state $\vert\alpha,\zeta\rangle$ is now obtained by
squeezing the vacuum and then displacing it:
\begin{equation}
\vert\alpha,\zeta\rangle = D(\alpha)S(\zeta)\vert 0 \rangle.
\end{equation}

\end{document}